\documentclass[a4paper,11pt]{article}

\usepackage[totalwidth=480pt,totalheight=680pt]{geometry}
\usepackage{rcs}
\RCS $Id: CDGMR07.tex,v 1.60 2008/05/29 07:49:42 cappe Exp $
\usepackage{amsmath,amssymb,amsfonts,amsthm,mathrsfs,bbm}
\usepackage{graphicx,color}
\usepackage{natbib}
\bibpunct{(}{)}{,}{a}{,}{,}
\usepackage{ifthen}
\newboolean{preview}
\setboolean{preview}{true}  
\ifthenelse{\not\boolean{preview}}{%
  \usepackage[notref,notcite]{showkeys}
  \usepackage{fancyheadings}
  \pagestyle{fancyplain}
  
  \lhead{\leftmark}
  \rhead{\thepage}
  \cfoot{{\tt \RCSId}}
  }%
{}

\newcommand{\bx}{\mathbf{x}}
\newcommand{\bX}{\mathbf{X}}
\newcommand{\by}{\mathbf{y}}
\newcommand{\E}{\mathbb{E}}

\newcommand{\1}{{\mathbbm{1}}}

\theoremstyle{definition}\newtheorem{algorithm}{Algorithm}

\begin{document}
\parskip 1ex 


\title{Adaptive Importance Sampling in General Mixture Classes\footnote{This work has been supported by
    the Agence Nationale de la Recherche (ANR)
    through the 2006-2008 project {\sf Adap'MC}. Both last authors are grateful
    to the participants to the BIRS {\sf 07w5079} meeting on ``Bioinformatics,
    Genetics and Stochastic Computation: Bridging the Gap'', Banff, for their
    comments on an earlier version of this paper. The last author also 
    acknowledges an helpful discussion with Geoff
    McLachlan. The authors wish to thank both referees for their encouraging comments.}}

\author{{\sc Olivier Capp\'e,}\\
  {\em LTCI, TELECOM ParisTech, CNRS}\\
  {\sc Randal Douc,}\\
  {\em TELECOM SudParis}\\
  {\sc Arnaud Guillin},\\
  {\em LATP, Ecole Centrale Marseille, CNRS}\\
  {\sc Jean-Michel Marin,}\\ 
  {\em Project \textsc{select}, INRIA Saclay, Universit\'{e} Paris Sud \&\ CREST, INSEE}\\
  \&\ {\sc Christian P. Robert}\\
  {\em CEREMADE, Universit\'e Paris Dauphine, CNRS \&\ CREST, INSEE}}
\date{}

\maketitle

\begin{abstract}
  In this paper, we propose an adaptive algorithm that iteratively updates both the weights and
  component parameters of a mixture importance sampling density so as to optimise the performance
  of importance sampling, as measured by an entropy criterion. The method, called M-PMC, is shown
  to be applicable to a wide class of importance sampling densities, which includes in particular
  mixtures of multivariate Student $t$ distributions. The performance of the proposed scheme is
  studied on both artificial and real examples, highlighting in particular the benefit of a novel
  Rao-Blackwellisation device which can be easily incorporated in the updating scheme.

\end{abstract}

\section{Introduction}
In recent years, there has been a renewed interest in using Monte Carlo procedures based on
Importance Sampling (abbreviated to IS in the following) for inference tasks. Compared to
alternatives such as Markov Chain Monte Carlo methods, the main appeal of IS procedures lies in the
possibility of developing parallel implementations, which becomes more and more important with the
generalisation of multiple core machines and computer clusters. Importance sampling procedures are 
also attractive in that they allow for an easy assessment of the Monte Carlo error 
(provided trustworthy estimates of the variance can be produced). 
As a consequence, it is therefore easier to construct learning mechanisms in IS settings
because of this ability to compare the errors. In many applications, the fact
that IS procedures may be tuned---by choosing an appropriate IS density---to minimise the
approximation error for a specific function of interest is also crucial. On the other hand, the
shortcomings of IS approaches are also well-known, including a poor scaling to highly multidimensional
problems and an acute sensitivity to the choice of the IS density combined with the fact that it
is impossible to come up with a universally efficient IS density. While there
exist a wide variety of solutions in the literature \citep[see, e.g.][Chapter
14]{Robert:Casella:2004}, this paper concentrates on the construction of adaptive 
importance sampling schemes in which the IS density is gradually improved based on the outcome of previous Monte Carlo draws.

While the method proposed here can be traced back to authors such as \cite{West:1992} or
\cite{Oh:Berger:1993}, it is closely related to the so-called Population Monte Carlo (henceforth
abbreviated to PMC) approach---in the sense of an iterated simulation of importance samples and in
opposition to Markov Chain Monte Carlo simulation that only produces a point at a time---introduced
by \cite{Cappe:Guillin:Marin:Robert:2004}. We briefly review the PMC approach, following the
exposition of \cite{Cappe:Guillin:Marin:Robert:2004} and
\cite{Douc:Guillin:Marin:Robert:2007a,Douc:Guillin:Marin:Robert:2007b}, in order to highlight the
differences with the present work. In PMC, a sample $(X_1,\dots,X_N)$ approximately distributed
from $\pi$, is repeatedly perturbed stochastically using an arbitrary Markov transition kernel
$q(x,x')$ so as to produce a new sample $(X_1',\dots,X'_N)$. Conducting a resampling step based on
the IS weights $\omega_i=\pi(X_i')/q(X_i,X_i')$, it is then possible to produce a new unweighted sample
$(\tilde X_1,\dots,\tilde X_N)$ that also constitutes an approximation to the target distribution
$\pi$.  Adaptivity in PMC was achieved by considering a transition kernel $q$ consisting of a
mixture of \emph{fixed transition kernels}
\begin{equation}
q_\alpha(x,x') = \sum_{d=1}^D\alpha_d q_d(x,x')\,, \qquad \sum_{d=1}^D\alpha_d = 1\,,
\label{eq:genericPMC}
\end{equation}
whose weights $\alpha_1,\dots,\alpha_D$ are tuned adaptively, along the iteration of the PMC
algorithm.  The adaptation procedure proposed by \cite{Douc:Guillin:Marin:Robert:2007a}, 
termed \emph{$D$-kernel PMC}, aims at
minimising the deviance or entropy criterion between the kernel $q_\alpha$ and the target $\pi$,
\begin{equation}
\label{eq:dev_genericPMC}
\mathfrak{E}(\pi,q_\alpha) = \mathbb{E}_{\pi}^X \left[ D(\pi\|q_\alpha(X, \cdot)) \right] \, ,
\end{equation}
where $D(p\|q) = \int \log\{p(x)/q(x)\}\, p(x) \text{d} x$ denotes the Kullback-Leibler divergence
(also called relative entropy), and where the expectation is 
taken under the target distribution $X\sim\pi$ since
the kernels $q_d(x,x')$ depend on the starting value $x$. 
In the sequel, we refer to the criterion in~\eqref{eq:dev_genericPMC} 
as the \emph{entropy criterion} since it is obviously related to the performance 
measure used in the cross-entropy method of \cite{rubinstein:kroese:2004}.
In \cite{Douc:Guillin:Marin:Robert:2007b},
a version of this algorithm was developed to minimise the asymptotic variance of the IS procedure,
\emph{for a specific function of interest}, in lieu of the entropy criterion.

A major limitation in the approaches of 
\cite{Douc:Guillin:Marin:Robert:2007a,Douc:Guillin:Marin:Robert:2007b} is that the proposal kernels
$q_d$ remain fixed over the iterative process while only the mixture weights $\alpha_d$ get 
improved. In the present contribution, we remove this limitation by extending the framework of 
\cite{Douc:Guillin:Marin:Robert:2007a} to allow for the adaption of IS densities of the form 
\begin{equation}\label{eq:paramo}
  q_{(\alpha,\theta)}(x)=\sum_{d=1}^D\alpha_d q_d(x;\theta_d)\,,
\end{equation}
with respect to \emph{both} the weights $\alpha_d$ and the internal parameters $\theta_d$ of the
component densities. The proposed updating mechanism is quite similar to the EM algorithm with the
E-step replaced by IS computations. As demonstrated through the example considered in
Section~\ref{sec:tee}, this adaptive scheme is applicable to very general families of latent-data
IS densities. A possible drawback of adapting the internal parameters $\theta_d$ of the component
densities is that it sometimes raises challenging robustness issues, particularly when
(multidimensional) scaling parameters are tuned. We thus propose a Rao-Blackwellisation
scheme that empirically appears to be very efficient while inducing a modest additional algorithmic
complexity.

Note again that we consider here the generic entropy criterion of \cite{Douc:Guillin:Marin:Robert:2007a}
rather than the function-specific variance minimisation objective of
\cite{Douc:Guillin:Marin:Robert:2007b}. This choice is motivated by the recognition that in most
applications, the IS density is expected to perform well for a range of typical functions of
interest rather than for a specific target function $h$. In addition, the generalisation of the
approach of \cite{Douc:Guillin:Marin:Robert:2007b} to a class of mixture IS densities that are
parameterised by more than the weights remains an open question (see also Section~\ref{sec:ConC}).
A second remark is that in contrast to the previously cited works and as
obvious in equation \eqref{eq:paramo}, we consider in this paper only
``global'' independent IS densities. 
Thus, the proposed scheme is based on genuine iterated importance sampling, contrary to what happens 
when using more general IS transition kernels as in~\eqref{eq:genericPMC}. Obviously, resorting to
moves that depend on the current sample is initially attractive because it allows 
for some local moves as opposed to the global
exploration required by independent IS densities. However, the fact that the entropy criterion
in~\eqref{eq:dev_genericPMC} is a global measure of fit tends to modify the parameters of each
transition kernel depending on its average performance over the whole sample, rather than locally. In
addition, structurally imposing a dependence on the points sampled at the previous iteration
induces some extra-variability which can be detrimental when more parameters are to be estimated.

The paper is organised as follows: In Section \ref{sec:norm},
we develop a generic updating scheme for independent IS mixtures \eqref{eq:paramo}, 
establishing that the integrated EM argument of \cite{Douc:Guillin:Marin:Robert:2007a} remains valid
in our setting. Note once again that the integrated EM update
mechanism we uncover in this paper is applicable to 
all missing data representations of the
proposal kernel, and not only to finite mixtures.
In Section \ref{sec:AnormA}, we consider the case of Gaussian mixtures which naturally extend the case of 
mixtures of Gaussian random walks with fixed covariance structure considered in 
\cite{Douc:Guillin:Marin:Robert:2007a,Douc:Guillin:Marin:Robert:2007b}. In Section \ref{sec:tee}, we 
show that the algorithm also applies to mixtures of multivariate $t$ distributions with the continuous 
scale mixing representation used in \cite{Peel:McLachlan:2000}.  Section \ref{sec:ConC} provides some 
conclusive remarks about the performance of this approach as well as possible extensions.

\section{Adapting the Importance Sampling Density}
\label{sec:norm}
\subsection{The M-PMC Algorithm}
When considering independent mixture IS densities of the form~\eqref{eq:paramo}, 
the entropy criterion $\mathfrak{E}$ defined in~\eqref{eq:dev_genericPMC} reduces 
to the Kullback-Leibler divergence between the target density $\pi$ and the 
mixture $q_{(\alpha,\theta)}$:
\begin{equation}\label{eq:div}
  \mathfrak{E}(\pi,q_{(\alpha,\theta)}) = D(\pi\|q_{(\alpha,\theta)}) 
   = \int\log\left(\frac{\pi(x)}{\sum_{
        d=1}^D\alpha_d q_d(x;\theta_d)}\right)\pi(x) \text{d} x \, .
\end{equation}
As usual in applications of the IS methodology to Bayesian inference, the target density $\pi$ is
known only up to a normalisation constant and we will focus on a self-normalised version 
of IS that solely requires the availability of an unnormalised version of
$\pi$ \citep{Geweke:1989}. As a side comment, note that while
$\mathfrak{E}(\pi,q_{(\alpha,\theta)})$ is a convex function of the weights
$\alpha_1,\dots,\alpha_D$ \citep{Douc:Guillin:Marin:Robert:2007a}, it generally fails to be so when also
optimising with respect to the component parameters $\theta_1,\dots,\theta_D$. Given that minimising \eqref{eq:div} in $(\alpha,\theta)$ is equivalent
to maximising
\begin{equation}
  \label{eq:maxent}
\int \log\left( \sum_{d=1}^D\alpha_d q_d(x; \theta_d) \right) \pi(x)
\,\text{d}x\,,  
\end{equation}
we are facing a task that formally resembles standard mixture maximum likelihood estimation 
but with an integration with respect to $\pi$ replacing the empirical sum over observations. 

This analogy suggests that it is possible to maximise the entropy criterion in~\eqref{eq:div} using
an approach based on the principle of the EM algorithm and, in particular, the use of the augmented
mixture representation (involving the indicator variables associated with each component of the
mixture). Before providing the details of the derivation in Section~\ref{sec:algo:details}, we
first state below the proposed adaptive IS algorithm which we will refer to as M-PMC (for
\emph{Mixture PMC}) in the following. Let $(X_{i,t})_{1 \leq i \leq N}$ and $(\alpha^{t,N},
\theta^{t,N})$ denote, respectively, the IS sample and the estimated mixture parameters at the
$t$-th iteration of the algorithm.

\begin{algorithm}
\label{alg}
{\sf (M-PMC Algorithm)
At iteration $t$,
\begin{enumerate}
\item Generate a sample $(X_{i,t})$ from the current mixture IS proposal \eqref{eq:paramo} 
parameterised by $(\alpha^{t,N}, \theta^{t,N})$ and compute the normalised importance weights
\begin{equation}
  \label{eq:mixtureISweights}
  \bar\omega_{i,t}=\frac{\pi(X_{i,t})}{\sum_{d=1}^D\alpha^{t,N}_dq_d(X_{i,t};\theta^{t,N}_d)}\bigg/
  \sum_{j=1}^N \frac{\pi(X_{j,t})}{\sum_{d=1}^D\alpha^{t,N}_dq_d(X_{j,t};\theta^{t,N}_d)}
\end{equation}
and the mixture posterior probabilities
\begin{equation}
  \label{eq:condprob-alg}
  \rho_d(X_{i,t};\alpha^{t,N},\theta^{t,N}) = {\alpha_d^{t,N}
  q_d(X_{i,t};\theta_d^{t,N})}\bigg/{\sum_{\ell=1}^D \alpha_\ell^{t,N} q_\ell(X_{i,t};\theta_\ell^{t,N})} \, , 
\end{equation}
for $i=1,\dots,N$ and $d=1,\dots,D$.
\item Update the parameters $\alpha$ and $\theta$ as
\begin{eqnarray}
\alpha^{t+1,N}_d &=& \sum_{i=1}^N\bar\omega_{i,t}\rho_d\left(X_{i,t};
	\alpha^{t,N},\theta^{t,N}\right)\,, \nonumber\\
\theta^{t+1,N}_d &=& \arg\max_{\theta_d}\left[\sum_{i=1}^N\bar\omega_{i,t}
 \rho_d\left(X_{i,t};\alpha^{t,N},\theta^{t,N}\right)
  \log\left\{q_d\left(X_{i,t};\theta^{t,N}_d\right)\right\}\right] \,,
  \label{eq:update_RB}
\end{eqnarray}
for $d=1,\dots,D$.
\end{enumerate}
}\end{algorithm}
The convergence of the algorithm may be monitored by computing the so-called normalised perplexity
$\exp(H^{t,N})/N$, where $H^{t,N} = -\sum_{i=1}^N \bar\omega_{i,t} \log \bar{\omega}_{i,t}$ is the
Shannon entropy of the normalised IS weights. The normalised perplexity provides an estimate of
$\exp [-\mathfrak{E}(\pi,q_{(\alpha^{t,N},\theta^{t,N})})]$ and, for sufficiently large $N$, it is non-decreasing with $t$.

\subsection{Detailed Derivation}
\label{sec:algo:details}

\subsubsection{Integrated Updates}
Starting from~\eqref{eq:maxent}, assume for the moment that integration with respect to $\pi$ is
feasible. In order to update the parameters of the independent IS density \eqref{eq:paramo}, we
will take advantage of the latent variable structure that underlines the objective
function~\eqref{eq:maxent}. The resulting algorithm---still theoretical at this stage as it
involves integration with respect to $\pi$---may be interpreted as an integrated EM
(Expectation-Maximisation) scheme that we now describe. Let
$\alpha^{t}=\left(\alpha_1^{t},\ldots,\alpha_D^{t}\right)$ and
$\theta^{t}=\left(\theta_1^{t},\ldots,\theta_D^{t}\right)$ denote, respectively, the mixture
weights and the component parameters at the $t$-th iteration of this integrated EM
algorithm.

As usual in mixtures, the latent variable $Z$ is the component indicator, 
with values in $\{1,\ldots,D\}$ such that the joint density $f$ of $x$ and $z$ satisfies
$$
f(z)=\alpha_z \quad \text{and} \quad f(x|z)=q_z(x;\theta_z) \, ,
$$
which produces \eqref{eq:paramo} as the marginal in $x$.
As in the standard EM algorithm, we can then take advantage of this latent variable representation.
Since the joint density of $X$ and $Z$ is $\alpha_z q_z(x; \theta_x)$, the expectation
corresponding to the E step of the EM algorithm is the expected complete log-likelihood, namely, at
iteration $t$ of our algorithm,
$$
\mathbb{E}_{\pi}^X \left[ \mathbb{E}_{(\alpha^t,\theta^t)}^Z \left\{ \log\left(
      \alpha_Z q_Z(X; \theta_Z) \right) | X \right\} \right]\,,
$$
where the inner expectation is computed under the conditional distribution of
$Z$ in the mixture model given the current value $(\alpha^t,\theta^t)$ of the parameters, i.e.
$$
f(z|x)=\alpha^t_z q_z(x;\theta_z^t) \bigg/ \sum_{d=1}^D \alpha^t_d
q_d(x;\theta_d^t)\,,
$$
while the outer expectation is under the distribution $X \sim \pi$.

The proposed updating mechanism then corresponds to setting the new
parameters $(\alpha^{t+1},\theta^{t+1})$ equal to
\begin{equation}
  \label{eq:ldsfk}
(\alpha^{t+1},\theta^{t+1})=\arg\max_{(\alpha,\theta)}\,
\E_\pi^X\left[\E_{(\alpha^t,\theta^t)}^Z\left\{\log(\alpha_Z q_Z(X;
    \theta_Z))|X\right\}\right]\,,
\end{equation}
as in the regular EM estimation of the parameters of a mixture, except for the extra expectation
over $X$. It is straightforward to check that the convexity argument used for the EM
algorithm also applies in this setup and, hence, that
$(\mathfrak{E}(\pi,q_{(\alpha^t,\theta^t)}))_{t\geq 1}$ is a non-decreasing sequence. Setting 
\begin{equation*}
  \rho_d(X;\alpha,\theta) = {\alpha_d
  q_d(X;\theta_d)}\bigg/{\sum_{\ell=1}^D \alpha_\ell q_\ell(X;\theta_\ell)} \, ,  
\end{equation*}
the maximisation program in~\eqref{eq:ldsfk} reduces to
\begin{align*}
 & \alpha^{t+1}=\arg\max_{\alpha}\E_\pi^X\left[\sum_{d=1}^D
  \rho_d(X;\alpha^t,\theta^t)\log(\alpha_d)\right]\,, \\
 & \theta^{t+1}=\arg\max_{\theta}\E_\pi^X\left[\sum_{d=1}^D
  \rho_d(X;\alpha^t,\theta^t)\log(q_d(X;\theta_d))\right]\,,
\end{align*}
where the first maximisation to be carried out under the constraint that $\sum_{d=1}^D \alpha^{t+1}_d = 1$. Hence,
\begin{align}
 & \alpha_d^{t+1}=\E_\pi^X\left[\rho_d(X;\alpha^t,\theta^t) \right]\,, \label{eq:updalpha} \\
 &  \theta_d^{t+1}=\arg\max_{\theta_d}\E_\pi^X\left[ 
  \rho_d(X;\alpha^t,\theta^t)\log(q_d(X;\theta_d)) \right]\,. \label{eq:updtheta}
\end{align}

As in the regular mixture estimation problem, the resolution of this
maximisation program ultimately depends on the shape of the density $q_d$. If $q_d$
belongs to an exponential family, it is easy to derive a closed-form solution for~\eqref{eq:updtheta}, which
however involves expectations under $\pi$.  Section \ref{sec:AnormA} provides an illustration
of this fact in the Gaussian case, while the non-exponential
Student's $t$ case is considered in Section \ref{sec:tee}.


\subsubsection{Approximate Updates}
To make the previous algorithm practical, adaptivity must be achieved by updating the parameters
based on the previously simulated IS sample. We thus start the algorithm by arbitrarily fixing the
mixture parameters $(\alpha^1,\theta^1)$ and we then sample from the resulting proposal
$\sum\alpha^1_d q_d(x;\theta^1_d)$ to obtain our initial sample $(X_{i,1})_{1\le i\le N}$,
associated with the latent variables $(Z_{i,1})_{1\le i\le N}$ that indicate from which component
of the mixture the corresponding $(X_{i,1})_{1\le i\le N}$ have been generated. From this stage, we
proceed recursively. Starting at iteration $t$ from a sample $(X_{i,t})_{1\le i\le N}$, associated
with the latent variables $(Z_{i,t})_{1\le i\le N}$ and the normalised IS weights
$(\bar\omega_{i,t})_{1\le i\le N}$ defined in~\eqref{eq:mixtureISweights}, we denote by
$(\alpha^{t+1,N},\theta^{t+1,N})$ the updated value of the mixture parameters.

To approximate~\eqref{eq:updalpha} and~\eqref{eq:updtheta}, \cite{Douc:Guillin:Marin:Robert:2007a} 
proposed the following update rule:
\begin{eqnarray}
\alpha^{t+1,N}_d & = & \sum_{i=1}^N\bar\omega_{i,t}\1\{Z_{i,t} = d\} \, , \nonumber \\
\theta^{t+1,N}_d & = & \arg\max_{\theta_d}\left[\sum_{i=1}^N\bar\omega_{i,t}
 \1\{Z_{i,t} = d\}
  \log\left\{q_d\left(X_{i,t};\theta^{t,N}_d\right)\right\}\right] \, .
  \label{eq:update_plain}
\end{eqnarray}
The computational cost of this update is of order $N$ whatever the number $D$ of components is,
since the weight and the parameter of each component are updated based only on the points that were
actually generated from this component. However, this observation also suggests
that~\eqref{eq:update_plain} may be highly variable when $N$ is small and/or $D$ becomes larger.
To make the update more robust, we here propose a simple Rao-Blackwellisation step that consists in
replacing $\1\{Z_{i,t} = d\}$ with its conditional expectation given $X_{i,t}$, that is,
$\rho_d\left(X_{i,t};\alpha^{t,N},\theta^{t,N}\right)$ defined in~\eqref{eq:condprob-alg}. The resulting parameters update is given by~\eqref{eq:update_RB}, which we selected for Algorithm~\ref{alg}.

Examining~\eqref{eq:condprob-alg} indicates that the evaluation of the posterior
probabilities $\rho_d (X_{i,t};\alpha^{t,N},\theta^{t,N})$ does not
represent a significant additional computation cost,
given that the denominator of this
expression has already been computed when evaluating the IS weights according
to~\eqref{eq:mixtureISweights}. The most significant difference between~\eqref{eq:update_RB} and~\eqref{eq:update_plain} is that, with the former,
all points contribute to the updating of the $d$-th component, for an overall
cost proportional to $D \times N$. Note however that in many applications of
interest, the most significant computational cost is associated with the
evaluation of $\pi$---which is performed exactly $N$ times per iteration---so 
that the cost of the update is mostly negligible, even
with the Rao-Blackwellised version.

\subsubsection{Convergence of the M-PMC Algorithm}
Convergence of the estimated parameters as $N$ increases can be established using the
same approach as in \cite{Douc:Guillin:Marin:Robert:2007a,Douc:Guillin:Marin:Robert:2007b}, 
relying mainly on the convergence property of triangular arrays of random variables (see Theorem A.1 in 
\citealp{Douc:Guillin:Marin:Robert:2007a}). For the Rao-Blackwellised version, assuming that for all $\theta$'s,
$\pi(q_d(\cdot;\theta_d)=0)=0$, for all $\alpha$'s and $\theta$'s,
$\rho_d(\cdot;\alpha,\theta)\log q_d(\cdot,\theta_d)\in L^1(\pi)$, and some 
(uniform in $x$) regularity conditions on $q_d(x;\theta)$ viewed as a function of $\theta$, 
yield
$$
\alpha^{t+1,N}_d \stackrel{\mathbb{P}}{\to}\alpha^{t+1}_d,
\qquad\qquad \theta^{t+1,N}_d \stackrel{\mathbb{P}}{\to}\theta^{t+1}_d
$$
when $N$ goes to infinity.
Note that we do not expand on the regularity conditions imposed on $q_d$ since, 
for the algorithm to be efficient, we definitely need a closed-form expression
on the parameter updates. It is then easier to
deal with the convergence of the approximation of these update formulas on a
case-by-case basis, as will be seen in the Gaussian example of Section~\ref{sec:AnormA}.

As a practical criterion for monitoring the convergence of the algorithm we recommend computing the
normalised perplexity $\exp(H^{t,N})/N$ (see Algorithm~\ref{alg}) and to interrupt adaptation when
it stabilises and/or becomes sufficiently close to 1. Note that in referring to $\exp(H^{t,N})$
(exponential of the Shannon entropy expressed in nat) as the \emph{perplexity}, we follow the
terminology in use in the field of natural language processing. The connection between the perplexity and the entropy criterion~(\ref{eq:dev_genericPMC}) is revealed by writing
\begin{equation}
  \exp \left[-\mathfrak{E}(\pi,q_{(\alpha,\theta)})\right] =
  \exp \left(\int - \log\frac{\pi_{\mathrm{unn}}(x)}{q_{(\alpha,\theta)}(x)}\pi(x) \text{d}x \right) 
  \left(\int \pi_{\mathrm{unn}}(x) \text{d}x\right) \, ,
  \label{eq:perp}
\end{equation}
where $\pi_{\mathrm{unn}}$ refers to the unnormalised version of $\pi$ which is effectively computable. Estimating the first integral in~\eqref{eq:perp} by self-normalised IS as 
$$
- \sum_{i=1}^N \bar\omega_{i,t} \log
\frac{\pi_{\mathrm{unn}}(X_{i,t})}{q_{(\alpha^{t,N},\theta^{t,N})}(X_{i,t})}
$$
and the second one by classical IS, as
$$
1/N\sum_{i=1}^N \pi_{\mathrm{unn}}(X_{i,t})/q_{(\alpha^{t,N},\theta^{t,N})}(X_{i,t}),
$$
indeed shows that $\exp(H^{t,N})/N$ is a consistent estimator of $\exp
[-\mathfrak{E}(\pi,q_{(\alpha^{t,N},\theta^{t,N})})]$. The entropy of the IS weights is
frequently used as a criterion for assessing the quality of an IS sample---together with the
so-called Effective Sample Size (ESS)
\citep{Chen:Liu:1996,Doucet:deFreitas:Gordon:2001,Cappe:Ryden:2004}. To the best of our knowledge,
however the strong connection between this criterion and the performance measure $\mathfrak{E}(\pi,q_{(\alpha^{t,N},\theta^{t,N})})$ used in the present work had not been noted before.

\subsubsection{Variance Estimation}
For the sake of completeness, we recall here the formula by which it is possible to estimate, from
the IS sample, the asymptotic variance of the IS estimate. If one considers a test function $h$ of
interest, the self-normalised IS estimation of its expectation under $\pi$ is $\widehat{\pi(h)} =
\sum_{i=1}^N \bar{\omega}_i h(X_i) $ and its asymptotic variance is given by
\[
  \upsilon(h) = \int \left\{ h(x)-\pi(h) \right\}^2 \pi^2(x)/q_{\alpha,\theta}(x) \text{d}x \, ,
\]
under the assumption that $\int (1+h^2(x)) \pi^2(x)/q_{\alpha,\theta}(x) \text{d}x < \infty$. The asymptotic variance $\upsilon(h)$ 
may thus be consistently estimated by $N \sum_{i=1}^N \bar\omega_i^2 \{ h(X_i)-\widehat{\pi(h)} \}^2$ \citep{Geweke:1989}. 


\section{The Gaussian mixture case}\label{sec:AnormA}
As a first example, we consider the case of $p$-dimensional Gaussian mixture IS densities of the form
$$
q_d(X;\theta_d)=\left\{\left(2\pi\right)^p|\Sigma_d|\right\}^{-1/2}
	\exp\left\{-\frac{1}{2}(X-\mu_d)^\text{T}\Sigma_d^{-1}(X-\mu_d)\right\}\,,
$$
where $\theta_d=\left(\mu_d,\Sigma_d\right)$ denotes the parameters of the $d$-th Gaussian component
density. This parametrisation of the IS density provides a general framework for approximating
multivariate targets $\pi$ and the corresponding algorithm is a straightforward instance
of the general framework discussed in the previous section.

\subsection{Update formulas}
\label{sec:gaus_update}
The integrated update formulas are obtained as the solution of
$$
\theta_d^{t+1,N}=\arg\min_{\theta}\E_\pi^X\left[\rho_d(X;\alpha^t,\theta^t)\left(\log|\Sigma_d|+(X-\mu_d)^\text{T}\Sigma_d^{-1}(X-\mu_d)\right)\right]\,.
$$
It is straightforward to check that the infimum is reached when, for $ d\in\{1,\ldots,D\}$,
$$
  \mu_d^{t+1}=\frac{\E_\pi^X\left[\rho_d(X;\alpha^t,\theta^t)X\right]}
  {\E_\pi^X\left[\rho_d(X;\alpha^t,\theta^t)\right]}\,,
$$
and
$$
  \Sigma_d^{t+1}=\frac{\E_\pi^X\left[\rho_d(X;\alpha^t,\theta^t)(X-\mu_d^{t+1})(X-\mu_d^{t+1})^\text{T}\right]}
  {\E_\pi^X\left[\rho_d(X;\alpha^t,\theta^t)\right]}\,.
$$

At iteration $t$ of the M-PMC algorithm, both the numerator
and the denominator of each of the above expressions are approximated using
self-normalised importance sampling.
Denoting $\1\{Z_{i,t} = d\}$ by $\xi_{i,t}$, the following empirical update
equations are obtained for the basic updating strategy (\ref{eq:update_plain}):
\begin{align} \alpha_d^{t+1,N} & = 
	\sum_{i=1}^N \bar\omega_{i,t} \xi_{i,t}\,, \nonumber \\
  \mu_d^{t+1,N} & = \frac{\sum_{i=1}^N \bar\omega_{i,t} \xi_{i,t} X_{i,t}} 
  { \sum_{i=1}^N \bar\omega_{i,t} \xi_{i,t}} = \sum_{i=1}^N \bar\omega_{i,t} \xi_{i,t}X_{i,t} \, \Big/ \alpha_d^{t+1,N} \, , \nonumber \\
  \Sigma_d^{t+1,N} & = \sum_{i=1}^N \bar\omega_{i,t} \xi_{i,t}(X_{i,t}-\mu_d^{t+1,N})(X_{i,t}
  -\mu_d^{t+1,N})^\text{T} \, \Big/ \alpha_d^{t+1,N}\,. \label{eq:gaussian_update}
\end{align}
For the Rao-Blackwellised scheme of Algorithm~\ref{alg}, the update is formally identical to the one above upon replacing $\xi_{i,t}$ by its conditional expectation
\begin{equation}
  \label{eq:gaussian_update_RB}
  \xi^{RB}_{i,t}=\rho_d(X_{i,t};\alpha^{t,N},\theta^{t,N}) \, .
\end{equation}
Note that, as discussed in Section~\ref{sec:algo:details}, establishing the convergence of
the parameter update in this 
Gaussian setting will only require the assumption that 
$\rho_d(x;\alpha,\theta)x^2$ is integrable with respect to $\pi$ \citep[see][]{Douc:Guillin:Marin:Robert:2007a}.


\subsection{A simulation experiment}
To illustrate the results of the algorithm presented above, we consider a toy
example in which the target density consists of a mixture of two
multivariate Gaussian densities. The appeal of this example is that it is
sufficiently simple to allow for an explicit characterisation of the attractive
points for the adaptive procedure, while still illustrating the variety of
situations found in more realistic applications. In particular, the model contains 
an attractive point that does not correspond to the global minimum of the
entropy criterion as well as some regions of attraction that can eventually
lead to a failure of the algorithm. The results obtained on this example also
illustrate the improvement brought by the Rao-Blackwellised update formulas
in~\eqref{eq:gaussian_update_RB}.

The target $\pi$ is a mixture of two $p$-dimensional Gaussian densities
such that
\[
\pi(x) = 0.5 \mathcal{N}(x; -s \mathbf{u}_p, \mathbf{I}_p) + 0.5 \mathcal{N}(x;
s \mathbf{u}_p, \mathbf{I}_p) \, ,
\]
when $\mathbf{u}_p$ is the $p$-dimensional vector whose coordinates are equal
to 1 and $\mathbf{I}_p$ stands for the identity matrix. In the sequel, we focus
on the case where $p=10$ and $s=2$. Note that one should not be misled by the
image given by the marginal densities of $\pi$: in the ten dimensional
space, the two components of $\pi$ are indeed very far from one another. It is
for instance straightforward to check that the Kullback-Leibler divergence
between the two components of $\pi$, $D\left\{\left.\mathcal{N}(s
    \mathbf{u}_p, \mathbf{I}_p) \right\| \mathcal{N}(-s \mathbf{u}_p,
  \mathbf{I}_p)\right\}$, is equal to $\frac12 \| 2 s \mathbf{u}_p \|^2 = 2 s^2
p$, that is 80 in the case under consideration. In particular, were we to
use one of the components of the mixture as an IS density for
the other, we know from the arguments exposed at the end of
Section~\ref{sec:norm} that the normalised perplexity
of the weights would eventually tend to $\exp(-80)$. This number is so
small that, for any feasible sample size, using one of the component
densities of $\pi$ as an IS instrumental density for the other component or
even for $\pi$ itself can only provide useless biased estimates.

The initial IS density $q_0$ is chosen here as the isotropic
ten-dimensional Gaussian density with a covariance matrix of
$5\mathbf{I}_p$. The performances of $q_0$ as an importance sampling
density, when compared to various other alternatives, are fully detailed in
Table~\ref{tab:ISperf} below but the general comment is that it corresponds to
a poor initial guess which would provide highly variable results when used with
any sample size under $50,000$.

\begin{table}[hbt]
  \centering
  \begin{tabular}{l|c|c|c} 
    Proposal & N-PERP & N-ESS & $\sigma^2(x_1)$ \\ \hline
      & & & \\
    $q_0$ $^\dagger$ & 6.5E-4 & 1.5E-4 & 37E3 \\
    Best fitting Gaussian  $^\dagger$ & 0.31 & 0.27 & 19 \\
    Target mixture  $^\dagger$ & 1  $^\dagger$ & 1  $^\dagger$ & 5  $^\dagger$ \\
    Best fitting Gaussian (defensive option) & 0.28 & 0.23 & 22 \\
    Best fitting two Gaussian mixture (defensive option) & 0.89 & 0.87 & 5.8 
  \end{tabular}
  \caption{Performance of various importance sampling densities in terms of N-PERP: 
  Normalised perplexity; N-ESS: Normalised Effective Sample Size; $\sigma^2(x_1)$: 
  Asymptotic variance of self-normalised IS estimator for the coordinate projection function
  $h(x)=x_1$. Quantities marked with a dagger sign are straightforward to determine, all others 
  have been obtained using IS with a sample size of one million.}
  \label{tab:ISperf}
\end{table}

In addition to figures related to the initial IS density
$q_0$, Table~\ref{tab:ISperf} also reports performance obtained with the best
fitting Gaussian IS density (with respect to the entropy criterion), which
is straightforwardly obtained as the centred Gaussian density
whose covariance matrix matches the one of $\pi$, that is, $\mathbf{I}_p + s^2
\mathbf{u}_p \mathbf{u}_p^{\text{T}}$. Of course the best possible performance
achievable with a mixture of two Gaussian densities, always with the
entropy criterion, is obtained when using $\pi$ as an IS density (second line of
Table~\ref{tab:ISperf}). Finally both final lines of Table~\ref{tab:ISperf}
report the best fit obtained with IS densities of the form $0.9
\sum_{d=1}^D \alpha_d \mathcal{N}(\mu_d, \Sigma_d) + 0.1 q_0(\cdot)$ when,
respectively, $D=1$ and $D=2$ (further comments on the use of these are given
below). As a general comment on Table~\ref{tab:ISperf}, note that the
variations of the perplexity of the IS weights, of the ESS and of the
asymptotic variance of the IS estimate for the coordinate projection function
are very correlated. This is a phenomenon that we have observed on many
examples and which justifies our postulate that minimising the entropy criterion does
provide very significant variance reductions for the IS estimate of ``typical''
functions of interest.

In this example, one may categorise the possible outcomes of adaptive IS
algorithms based on mixtures of Gaussian IS densities into mostly four
situations:
\begin{description}
\item[Disastrous (D.)] After $T$ iterations of the M-PMC scheme, $q_{(\alpha^T,\theta^T)}$ is not a
  valid IS density (in the sense that the importance sampling unbiasedness property does not hold
  due to support restrictions) and may lead to inconsistent estimates. Typically, this
  may happen if $q_{(\alpha^T,\theta^T)}$ becomes much too peaky with light
  tails. As discussed above, it will also practically be the case if the
  algorithm only succeeds in fitting $q_{(\alpha^T,\theta^T)}$ to one of both
  Gaussian modes of $\pi$. Another disastrous outcome is when the direct
  application of the adaptation rules described above leads to numerical problems,
  usually due to the poor conditioning of some of the covariance matrices $\Sigma_d$.
  Rather than fixing these issues by ad-hoc solutions (eg. diagonal loading), which
  could nonetheless be useful in practical applications, we consider below more
  principled ways of making the algorithm more resistant to such failures.
\item[Mediocre (M.)] After adaptation, $q_{(\alpha^T,\theta^T)}$ is not
  significantly better than $q_0$ in terms of the performance criteria
  displayed in Table~\ref{tab:ISperf} and, in this case, the adaptation is
  useless.
\item[Good (G.)] After $T$ iterations, $q_{(\alpha^T,\theta^T)}$ selects the best
  fitting Gaussian approximation (second line of Table~\ref{tab:ISperf}) which
  already provides a very substantial improvement as it results in variance
  reductions by about four orders of magnitude for typical functions of
  interest.
\item[Excellent (E.)] After $T$ iterations, $q_{(\alpha^T,\theta^T)}$ selects the
  best fitting mixture of two Gaussian densities, which in this somewhat
  artificial example corresponds to a perfect fit of $\pi$. Note, however that,
  the actual gain over the previous outcome is rather moderate with a reduction
  of variance by a factor less than four.
\end{description}
Of course, a very important parameter here is the IS sample size $N$: for a
given initial IS density $q_0$, if $N$ is too small, any method based on
IS is bound to fail, conversely when $N$ gets large all reasonable algorithms
are expected to reach either the G. or E. result. Note that with local
adaptive rules such as the ones proposed in this paper, it is not possible to
guarantee that only the E. outcome will be achieved as the best fitting
Gaussian IS density is indeed a stationary point (and in fact a local
minimum) of the entropy criterion. So, depending on the initialisation,
there always is a non zero probability that the algorithm converges to the
G. situation only.

To focus on situations where algorithmic robustness is an issue, we purposely
chose to select a rather small IS sample size of $N=5,000$ points. As discussed
above, direct IS estimates using $q_0$ as IS density would be
mostly useless with such a modest sample size. We evaluated four algorithmic
versions of the M-PMC algorithm. The first, \emph{Plain M-PMC}, uses the parameter update formulas
in~\eqref{eq:gaussian_update} and $q_0$
is only used as an initialisation value, which is common to all $D$ components
of the mixture (which also initially have equal weights). Only the means of the
components are slightly perturbed to make it possible for the adaptation
procedure to actually provide distinct mixture components.  One drawback of the
plain M-PMC approach is that we do not ensure during the course of the algorithm
that the adapted mixture IS density remains appropriate for IS approximations, 
in particular that it provides reliable estimates of the parameter update formulas. 
To guarantee that the IS weights stay well behaved, we consider a version of the M-PMC 
algorithm in which the IS density is of the form
\[
(1-\alpha_0) \sum_{d=1}^D \alpha_d \mathcal{N}(\mu_d, \Sigma_d) + \alpha_0 q_0
\]
with the difference that $\alpha_0$ is a fixed parameter which is not adapted. 
The aim of this version, which we call \emph{Defensive M-PMC} in reference to the work of
\cite{Hesterberg:1995}, is to guarantee that the importance function remains
bounded by $\alpha_0^{-1} \pi(x)/q_0(x)$, whatever happens during the
adaptation, thus guaranteeing a finite variance. 
Since $q_0$ is a poor IS density, it is preferable to keep
$\alpha_0$ as low as possible and we used $\alpha_0 = 0.1$ in all the following
simulations. As detailed in both last lines of Table~\ref{tab:ISperf}, this
modification will typically slightly limit the performances achievable by the
adaptation procedure, although this drawback could probably be avoided by
allowing for a decrease of $\alpha_0$ during the iterations of the M-PMC. The
parameter update formulas for this modified mixture model are very easily
deduced from~\eqref{eq:gaussian_update}
and are omitted here for the sake of conciseness. The third version we
considered is termed \emph{Rao-Blackwellised M-PMC} and consists in replacing the
update
equations~\eqref{eq:gaussian_update} by
their Rao-Blackwellised
version~\eqref{eq:gaussian_update_RB}.
Finally, we consider a fourth option in which both the defensive mixture
density and the Rao-Blackwellised update formulas are used.

All simulations were carried out using a sample size of $N=5,000$, 20
iterations of the M-PMC algorithm and Gaussian mixtures with $D=3$ components.
Note that we purposely avoided to chose $D=2$ to avoid the very artificial ``perfect fit''
phenomenon. This also means that for most runs of the algorithm, at least one
component will disappear (by convergence of its weight to zero) or will be
duplicated, with several components sharing very similar parameters.

\begin{table}[hbt]
  \centering
  \begin{tabular}{l|c|c|c|c} 
    & Disastrous & Mediocre & Good & Excellent \\ \hline
    Plain & 55 & 0 & 33 & 12 \\ \hline
    Defensive & 13 & 51 & 30 & 6 \\ \hline
    R.-B. & 18 & 1 & 70 & 11 \\ \hline
    Defensive $+$ R.-B. & 5 & 11 & 76 & 8 
  \end{tabular}
  \caption{Number of outcomes of each category for the four algorithmic versions, as recorded from 100 independent runs.}
  \label{tab:Gaus_perf}
\end{table}

Table~\ref{tab:Gaus_perf} display the performance of the four algorithms in repeated independent
adaptation runs. The most significant observation about Table~\ref{tab:Gaus_perf} is the large gap
in robustness between the non Rao-Blackwellised versions of the algorithm, which returned disastrous
or mediocre results in about 60\% of the cases, a fraction that falls bellow 20\% when the
Rao-Blackwellised update formulas are used. Obviously the fact that the Rao-Blackwellised updates
are based on all simulated values and not just on those actually simulated from a particular mixture
component is a major source of robustness of the method when the sample size $N$ is small, given
the misfit of the initial IS density $q_0$. The same remark also applies when the M-PMC algorithm
is to be implemented with a large number $D$ of components. The role of the defensive mixture component
is more modest although it does improve the performance of both versions of the algorithm (non
Rao-Blackwellised and Rao-Blackwellised altogether), at the price of a slight reduction of the frequency
of the ``Excellent'' outcome. Also notice that the results obtained when the defensive mixture
component is used are slightly beyond those of the unconstrained adaptation (see
Table~\ref{tab:ISperf}). The frequency of the perfect or ``Excellent'' match is about 10\% for all
methods but this is a consequence of the local nature of the adaptation rule as well as of the
choice of the initialisation of the algorithm. It should be stressed however that as we are not
interested in modelling $\pi$ by a mixture but rather that we are seeking good IS densities, 
the solutions obtained in the G. or E. situations are only mildly different in this respect (see
Table~\ref{tab:ISperf}). As a final comment, recall that the results presented above have been 
obtained with a fairly small sample size of $N=5,000$. Increasing $N$ quickly reduces the failure 
rate of all algorithms: for $N=20,000$ for instance, the failure rate of the plain M-PMC algorithm 
drops to 7/100 while the Rao-Blackwellised versions achieve either the G. or E. result (and 
mostly the G. one, given the chosen initialisation) for all runs.

\section{Robustification via mixtures of multivariate $t$'s}\label{sec:tee}
We now consider the setting of a proposal composed of a mixture of $p$-dimensional $t$ distributions,
\begin{equation}\label{eq:mix'o't}
\sum_{d=1}^D \alpha_d \mathcal{T}(\nu_d,\mu_d,\Sigma_d)\, .
\end{equation}
We here follow the recommendations of \cite{West:1992} and
\cite{Oh:Berger:1993} who proposed using mixtures of $t$ 
distributions in importance sampling. The $t$ mixture is 
preferable to a normal mixture because of its heavier 
tails that can capture a wider range of non-Gaussian 
targets with a smaller number of components. This
alternative setting is more challenging however 
and one must take advantage of the missing variable representation
of the $t$ distribution itself to achieve a closed-form updating of the parameters
$(\mu_d,\Sigma_d)_d$ approximating \eqref{eq:updtheta}, since a true closed-form 
cannot be derived.

\subsection{The latent-data framework}\label{sub:com}
Using the classical normal/chi-squared decomposition of the $t$ distribution, a joint
distribution associated with the $t$ mixture proposal \eqref{eq:mix'o't} is
\begin{align*}
  f(x,y,z) &\propto \alpha_z |\Sigma_z|^{-1/2} \exp\left\{ -(x-\mu_z)^\text{T}
    \Sigma_z^{-1} (x-\mu_z) y / 2 \nu_z \right\} y^{(\nu_z+p)/2-1} e^{-y/2}\\
  &\propto \alpha_z \,\varphi(x;\mu_z,\nu_z \Sigma_z /y) \,
  \varsigma(y;\nu_z/2,1/2)\,,
\end{align*}
where, as above, $x$ corresponds to the observable in \eqref{eq:mix'o't},
$z$ corresponds to the mixture indicator, and $y$ corresponds
to the $\chi^2_\nu$ completion.  The
normal density is denoted by $\varphi$ and the gamma density by $\varsigma$.
Both $y$ and $z$ correspond to latent variables in that the integral of the above in
$(y,z)$ returns \eqref{eq:mix'o't}.

In the associated M-PMC algorithm, we only update the expectations and the covariance structures 
of the $t$ distributions and not the number of degrees of freedom, given that there is
no closed-form solution for the later. In that case,
$\theta_d=(\mu_d,\Sigma_d)$ and, for each $d=1,\ldots,D$, the number of degrees
of freedom $\nu_d$ is fixed.
At iteration $t$, the integrated EM update of the parameter will involve the
following ``E'' function
$$
Q\{ (\alpha^t,\theta^t),(\alpha,\theta)\} =
\E^X_\pi\left[\E^{Y,Z}_{(\alpha^t,\theta^t)}\left\{\left.  \log(\alpha_Z) +
      \log(\varphi(X;\mu_Z,\nu_Z \Sigma_z /Y))\right|X \right\}\right]\,,
$$
since the $\chi^2$ part does not involve the parameter
$\theta=(\mu,\Sigma)$.  Given that
$$
Y,Z|X,\theta \sim f(y,z|x)\propto \alpha_z \,\varphi(x;\mu_z,\nu_z \Sigma_z /y)
\, \varsigma(y;\nu_z/2,1/2)\,,
$$
we have that
$$
Y|X,Z=d,\theta\sim \mathcal{G}a\left[ (\nu_d+p)/2, \frac{1}{2}\left\{
    1+(X-\mu_d)^\text{T} \Sigma_d^{-1} (X-\mu_d)/\nu_d \right\}\right]
$$
and therefore
\begin{align*}
  Q\{ (\alpha^t,\theta^t),(\alpha,\theta)\} 
  &= \E^X_{\pi}\left[ \sum_{d=1}^D \rho_d(X;\alpha^t,\theta^t) \log(\alpha^\prime_d) \right] \\
  & \qquad - \frac{1}{2}\,\E^X_{\pi}\Bigg[\sum_{d=1}^D
    \rho_d(X;\alpha^t,\theta^t)\bigg\{ \log |\Sigma_d|+
    (X-\mu_d)^\text{T} {\Sigma_d}^{-1} (X-\mu_d)  \\
  & \qquad \qquad \times \frac{\nu_d+p}{\nu_d+(X-\mu_d^t)^\text{T}
        (\Sigma_d^t)^{-1} (X-\mu_d^t)} \bigg\} \Bigg] \,,
\end{align*}
where we have used both the notation,
$$
\rho_d(X;\alpha^t,\theta^t) = \mathbb{P}_{\alpha^t,\theta^t}(Z=d|X) = \frac{\alpha_d^t
  t(x;\nu_d,\mu_d^t,\Sigma_d^t)}{\sum_{\ell=1}^D \alpha_\ell^t
  t(x;\nu_\ell,\mu_\ell^t,\Sigma_\ell^t)}  \, ,
$$
with $t(x;\nu,\mu,\Sigma)$ denoting the $\mathcal{T}(\nu,\mu,\Sigma)$ density, and the fact that
$$
 \gamma_d(X;\theta^t) = \E^Y_{\theta^t}\left\{Y/\nu_d|X,Z=d\right\} =
\frac{\nu_d+p}{\nu_d+(X-\mu_d^t)^\text{T} (\Sigma_d^t)^{-1} (X-\mu_d^t)} \,.
$$
Therefore, the ``M'' step of the integrated EM update is
\begin{eqnarray*}
  \alpha^{t+1}_d &=& \E^X_{\pi}\left[\rho_d(X;\alpha^t,\theta^t) \right] \, , \\
  \mu^{t+1}_d &=& \frac{\E^X_{\pi}\left[ \rho_d(X;\alpha^t,\theta^t) 
      \gamma_d(X;\theta^t) X \right]}
  {\E^X_{\pi}\left[ \rho_d(X;\alpha^t,\theta^t)
      \gamma_d(X;\theta^t) \right]} \, , \\
  \Sigma^{t+1}_d &=& \frac{\E^X_{\pi}\left[ \rho_d(X;\alpha^t,\theta^t)
      \gamma_d(X;\theta^t)
      (X-\mu^{t+1}_d)(X-\mu^{t+1}_d)^\text{T} \right]}
  {\E^X_{\pi}\left[ \rho_d(X;\alpha^t,\theta^t) \right]}\,.
\end{eqnarray*}
While the first update is the generic weight modification \eqref{eq:updalpha},
the latter formulae are (up to the integration with respect to $X$) essentially
those found in \cite{Peel:McLachlan:2000} for a mixture of $t$ distributions. 

\subsection{Parameter update}\label{sub:updaT}
As in Section~\ref{sec:gaus_update}, the empirical update equations are obtained by using self-normalised IS with weights $\bar{\omega}_{i,t}$ 
given by \eqref{eq:mixtureISweights} for both
the numerator and the denominator of each of the above expressions. The Rao-Blackwellised approximation based on~(\ref{eq:update_RB}) yields
\begin{align*}
  \alpha^{t+1,N}_d &= \sum_{i=1}^N \bar{\omega}_{i,t}\,\rho_d(X_{i,t};\alpha^{t,N},\theta^{t,N})\,, \nonumber \\
  \mu^{t+1,N}_d & = \frac{\sum_{i=1}^N \bar{\omega}_{i,t} \,\rho_d(X_{i,t};\alpha^{t,N},\theta^{t,N}) \, \gamma_d(X_{i,t};\theta^{t,N})
    \, X_{i,t}}{\sum_{i=1}^N \bar{\omega}_{i,t} \,
    \rho_d(X_{i,t};\alpha^{t,N},\theta^{t,N}) \, \gamma_d(X_{i,t};\theta^{t,N}) } \,, \nonumber \\
  \Sigma^{t+1,N}_d &= \frac{\sum_{i=1}^N \bar{\omega}_{i,t} \,\rho_d(X_{i,t};\alpha^{t,N},\theta^{t,N}) \, \gamma_d(X_{i,t};\theta^{t,N}) \,
    (X_{i,t}-\mu^{t+1,N}_d)(X_{i,t}-\mu^{t+1,N}_d)^\text{T}}{\sum_{i=1}^N
    \bar{\omega}_{i,t} \, \rho_d(X_{i,t};\alpha^{t,N},\theta^{t,N})}\, ,
\end{align*}
while the standard update equations, based on~(\ref{eq:update_plain}), 
are obtained by replacing $\rho_d(X_{i,t};\alpha^{t,N},\theta^{t,N})$ 
by $\1\{X_{i,t}=d\}$ in the above equations.

\subsection{Pima Indian example}
As a realistic if artificial illustration of the performances of the $t$ mixture \eqref{eq:mix'o't},
we study the posterior distribution of the parameters of a probit model.
The corresponding dataset is borrowed from the {\sf MASS} library of {\sf R}
\citep{CRAN}. It consists in
the records of 532 Pima Indian women who were tested by the U.S.~National
Institute of Diabetes and Digestive and Kidney Diseases for diabetes. Four
quantitative covariates were recorded, along with the presence or absence of
diabetes. The corresponding probit model analyses the presence of diabetes, i.e.~
$$
\mathbb{P}_\beta(y=1|\bx) = 1 - \mathbb{P}_\beta(y=0|\bx) = \Phi(\beta_0+\bx^\text{T}(\beta_1,\beta_2,\beta_3,\beta_4))
$$
with $\beta=(\beta_0,\ldots,\beta_4)$, $\bx$ made of four covariates, the number of pregnancies, the plasma glucose
concentration, the body mass index weight in kg/(height in m)$^2$, and the age, and
$\Phi$ corresponds to the cumulative distribution function of the standard normal.
We use the flat prior distribution $\pi(\beta|\bX)\propto 1$; in that case, the $5$-dimensional target
posterior distribution is such that
$$
\pi(\beta|\by,\bX)\propto \prod_{i=1}^{532}\left[\Phi\{\beta_0+(\bx^{i})^\text{T}
(\beta_1,\beta_2,\beta_3,\beta_4)\}\right]^{y_i}\left[1-\Phi\{\beta_0+(\bx^{i})^\text{T}
(\beta_1,\beta_2,\beta_3,\beta_4)\}\right[^{1-y_i}
$$
where $\bx^i$ is the value of the covariates for the $i$-th individuals and $y_i$ is the response of the $i$-th individuals.

\begin{figure}[p] \centering
  \includegraphics[keepaspectratio,width=5cm]{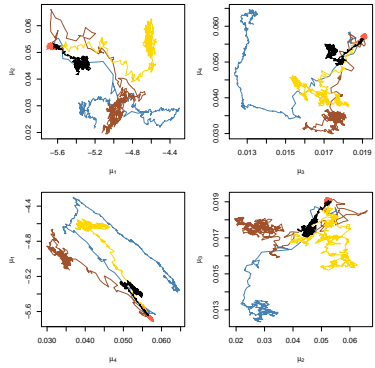}
  \caption{{\bf Pima Indians:} Evolution of the components of the five
    $\mu_d$'s over $500$ iterations plotted by pairs: {\em (clockwise from
      upper left side)} $(1,2)$, $(3,4)$, $(4,1)$ and $(2,3)$. The colour code
    is blue for $\mu_1$, yellow for $\mu_2$, brown for $\mu_3$ and red for
    $\mu_4$. The additional dark path corresponds to the estimate of $\beta$.
    All $\mu_d$'s were started in the vicinity of the MLE $\hat\beta$.
    \label{fig:pim1}}
\end{figure}

\begin{figure}[p] \centering
  \includegraphics[keepaspectratio,width=5cm]{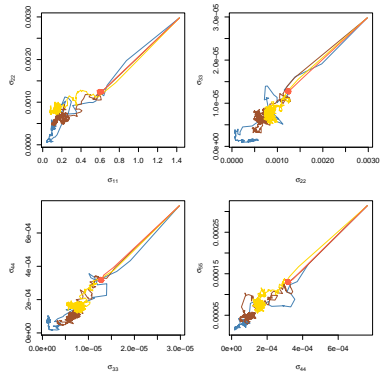}
  \caption{{\bf Pima Indians:} Evolution of the five $\Sigma_d$'s over $500$
    iterations plotted by pairs for the diagonal elements: {\em (clockwise from
      upper left side)} $(1,2)$, $(3,4)$, $(4,1)$ and $(2,3)$. The colour code
    is blue for $\Sigma_1$, yellow for $\Sigma_2$, brown for $\Sigma_3$ and red
    for $\Sigma_4$.  All $\Sigma_d$'s were started at the covariance matrix of
    $\hat\beta$ produced by {\tt R glm()} procedure.
    \label{fig:pim2}}
\end{figure}

\begin{figure}[p] \centering
  \includegraphics[keepaspectratio,width=5cm]{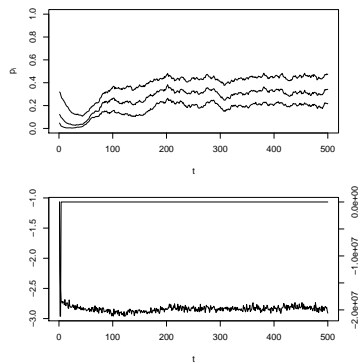}
  \caption{{\bf Pima Indians:} Evolution of the cumulated weights {\em (top)} and of the estimated
    entropy divergence $\mathbb{E}^\pi[\log(q_{\alpha,\theta}(\beta))]$  {\em (bottom)}.
    \label{fig:pim3}}
\end{figure}

We first present some results for $N=10,000$ sample points and $T=500$ iterations on Figures
\ref{fig:pim1}---\ref{fig:pim3}, based on a mixture with $4$ components
and with the degrees of freedom chosen as $\nu=(3,6,9,18)$, respectively, when
using the non Rao-Blackwellised version \eqref{eq:update_plain}. 
The unrealistic value of $T$ is chosen purposely to illustrate the lack of 
stability of the update strategy when not using the Rao-Blackwellised version.
Indeed, as can be seen from Figure \ref{fig:pim1}, which describes the
evolution of the $\mu_d$'s, some components vary quite widely over iterations,
but they also correspond to a rather stable overall estimate of $\beta$,
$\sum_{i=1}^N \bar\omega_{i,T} \beta^{i,T}$,
equal to $(-5.54,0.051,0.019,0.055,0.022)$ over most iterations.  When
looking at Figure \ref{fig:pim3}, the quasi-constant entropy estimate after
iteration 100 or so shows that, even in this situation, there is little 
need to perpetuate the iterations till the $500$-th.

\begin{figure}[p] \centering
  \includegraphics[keepaspectratio,width=5cm]{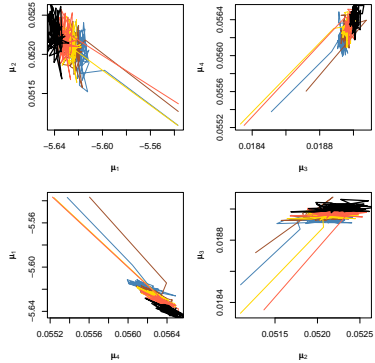}
  \caption{{\bf Pima Indians:} Evolution of the components of the five
    $\mu_d$'s over $50$ Rao-Blackwellised iterations plotted by pairs: {\em (clockwise from
      upper left side)} $(1,2)$, $(3,4)$, $(4,1)$ and $(2,3)$. The colour code
    is blue for $\mu_1$, yellow for $\mu_2$, brown for $\mu_3$ and red for
    $\mu_4$. The additional dark path corresponds to the estimate of $\beta$.
    All $\mu_d$'s were started in the vicinity of the MLE $\hat\beta$.
    \label{fig:pimRB1}}
\end{figure}

\begin{figure}[p] \centering
  \includegraphics[keepaspectratio,width=5cm]{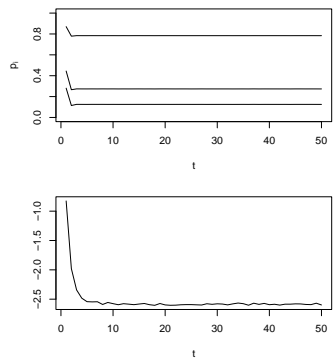}
  \caption{{\bf Pima Indians:} Evolution of the cumulated weights {\em (top)}
    and of the estimated entropy divergence 
    $\mathbb{E}^\pi[\log(q_{\alpha,\theta}(\beta))]$ {\em (bottom)} for the Rao-Blackwellised version.
    \label{fig:pimRB2}}
\end{figure}

Using a Rao-Blackwellised version of the updates shows a strong stabilisation for the updates
of the parameters $\alpha_d$ and $(\mu_d,\Sigma_d)$, both in the number of iterations and in
the range of the parameters. The approximation to the Bayes estimate is obviously very close 
to the above estimation $(-5.63,0.052,0.019,0.056,0.022)$. Figures \ref{fig:pimRB1} and \ref{fig:pimRB2}
show the immediate stabilisation provided by the Rao-Blackwellisation step.
In this example, which is quite typical in this respect, we recommend to use less than $T=10$
iterations in order to reserve most of the computational effort for increasing $N$, which is
essential during the first adaptation steps (because the initial IS density is poor) and for the
accuracy of the IS approximation in the final steps of the algorithm. Comparing the plain and Rao-Blackwellised update formulas, will really depend on how costly the parameter update is---and thus on the dimension of the model---compared to the other computational costs, and in particular the evaluation of the likelihood, which mostly depends on the number of observations. In the present case, the increase in run-time due to the use the Rao-Blackwellised formulas instead of the plain ones is negligible.

\section{Conclusions}\label{sec:ConC}
The M-PMC algorithm provides a flexible and robust framework for adapting
general importance sampling densities represented as mixtures. The extension to
mixtures of $t$ distribution broadens the scope of the method by allowing
approximation of heavier tail targets. Moreover, we can extend here the
remarks made in \cite{Douc:Guillin:Marin:Robert:2007a,
  Douc:Guillin:Marin:Robert:2007b}, namely that the update mechanism provides
an early stabilisation of the parameters of the mixture. It is therefore
unnecessary to rely on a large value of $T$: with large enough sample sizes $N$
at each iteration---especially on the initial iteration that requires many
points to counter-weight a potentially poor initial proposal---, it is quite
uncommon to fail to spot a stabilisation of both the estimates and of the
entropy criterion within a few iterations.

While this paper relies on the generic entropy criterion to update the mixture
density, we want to stress that it is also possible to use a more focussed
deviance criterion, namely the $h$-entropy
\begin{equation*}
  \mathfrak{E}_h(\pi,q_{(\alpha,\theta)}) = D( \pi_h \| q_{(\alpha,\theta)} ) \, ,
\end{equation*}
with
$$
\pi_h(x) \propto |h(x)-\pi(h)|\pi(x)\,,
$$
that is tuned to the estimation of a particular function $h$, as it is
well-known that the optimal choice of the importance density for the
self-normalised importance sampling estimator is exactly $\pi_h$. Since the
normalising constant in $\pi_h$ does not need to be known, one can derive an
adaptive algorithm which resembles the method presented in this paper.
It is expected that this modification will be helpful in reaching IS densities that provide a low approximation error for a specific function $h$, which is also a desirable feature of importance sampling in several applications.

\bibliography{bib/biblio}

\end{document}